\documentstyle[12pt]{article}
\begin{document}

 \def\Bbb{\bf }

\title{Relativizing relativity}
\author{K. Svozil\\
 {\small Institut f\"ur Theoretische Physik,}
  {\small Technische Universit\"at Wien }     \\
  {\small Wiedner Hauptstra\ss e 8-10/136,}
  {\small A-1040 Vienna, Austria   }            \\
  {\small e-mail: svozil@tuwien.ac.at}\\
  {\small www: http://tph.tuwien.ac.at/$\widetilde{\;\;}\,$svozil}}
\date{ }
\maketitle

\begin{flushright}
{\scriptsize http://tph.tuwien.ac.at/$\widetilde{\;\;}\,$svozil/publ/relrel.$\{$ps,tex$\}$}
\end{flushright}

\begin{abstract}
Special relativity theory is generalized to two or more
``maximal'' signalling speeds.
This framework is discussed in three contexts:
(i) as a scenario for superluminal signalling and motion,
(ii) as the possibility of two or more ``light'' cones due to
the a ``birefringent'' vaccum, and
(iii) as a further extension of conventionality beyond synchrony.
\end{abstract}

%0
%
%Relativity theory, particularly the special theory, is a fantastically
%successful theory. It is
%tempting to believe that relativity theory
%is one of the eternal truths that will be with us as science
%progresses, just like elementary arithmetic and the infinity of the
%set of prime numbers. Moreover, relativity theory is an archetype of a
%Pythagorean theory: it is all beauty, symmetry and clarity. It seems
%remote
%from the ``dirty roads'' of the physical practices and in almost perfect
%accord with and almost  above all experimental evidence,
%just like Newtonian concepts of space and time one hundred years
%ago.
%
%The amazing consistency of relativity theory and quantum theory has been
%subsumed into the expression of a ``peaceful coexistence''
%\cite{shimony2,shimony3}.
%Relativity theory therefore does not need any further commitment from
%the author and the reader besides possibly assurances that they do
%not belong to those
%weird folks pursuing wacky thoughts of ``disproofing relativity.''
%
%Finally, in physics only experiments and the phenomena count.
%History tells us that many of our would-be heavens of eternal harmony
%have collapsed, leaving only thin air behind.

\section{General framework}
In what follows, we shall study {\em two}  signal types with
two different signal velocities generating
two different sets of Lorentz frames associated with two types of
``light'' cones.
(A generalization to an arbitrary number of signals is straightforward.)
This may seem implausible and even misleading at first,
since
from two different ``maximal'' signal velocities only one can be truly
maximal. Only the maximal one appears to be the natural candidate for
the generation of Lorentz frames.

However,  it may sometimes be physically reasonable to consider
frames obtained by nonmaximal speed signalling.
What could such {\em subluminal coordinates}, as they
may be called, be good for?

(i) First of all, they may be useful for intermediate description levels
\cite{anderson:73,schweber:93} of physical theory. These description
levels may either be irreducible or derivable from some more
foundamental level. Such considerations appear to be closely related to
system science.

(ii)
By analogy, we may also consider faster-than-light ``signalling'' generating
{\em superluminal coordinates} \cite{recami:85,sush-86}.
Presently, faster-than-light ``signalling'' can, for instance, be
realized
by superluminal charge-current patterns; e.g., by the coordinated motion
of aggregates of electrically charged particles
\cite{bolo-ginz:72,arda:84}.
``Signals'' of the above type cannot convey
useful information and therefore cannot possibly be utilized to violate causality.
But it could also be speculated that
in the distant future signals of yet unknown type might be discovered
which travel faster than light. In this view, the ``second''
type of ``light'' cone just has not been discovered yet;  its
discovery being independent of and beyond the scope of these
considerations.

(iii) Thirdly, the standard debate of conventionality in relativity
theory which concentrates on synchrony can be extended to
arbitrary signalling speeds as well. This amounts to a splitting of
relativity theory into a section dealing with operational meaningful
conventions and another section expressing the physical content, in
particular covariance;
i.e., the form invariance of the equations of motion under the resulting
space-time transformations.

In all these cases the following considerations may yield a clearer
understanding of seemingly
``paradoxical'' effects such as time travel
\cite{godel-rmp,recami:87,nahin}.
 Thus
it may  not appear totally unreasonable to consider
generalized system
representations in which more than one signalling speeds are used to
generate space and time scales. The transformation properties
of such scales are then defined
{\em relative to the signal} invoked.

\section{Extending conventionality}

One of the greatest achievements of Einstein's theory of relativity is
the operational approach to space and time:
Already in Einstein's original article \cite{ein-05},
space as well as time scales
are generated by physical procedures and observables which are based upon
empirical phenomena which and on intrinsically meaningful concepts
\cite{bos,toffoli:79,svo5,svo-86,roessler-87,roessler-92,svozil-93}.
Such a requirement is by no means trivial.
For instance,  different  description levels may
use different signals (e.g., sound, waves of any form, light, $\ldots$).

Thereby, certain conventions have to be assumed,
which again have an operational meaning by refering to purely physical terms.
For instance, standard synchrony
at spatially separated locations is conventionalized
by ``radar procedures;'' i.e., by sending a signal back and forth
between two spatially separated clocks.
The conventionality of synchrony has been discussed, among others, by
Reichenbach \cite{reich-58},
Gr\"unbaum \cite{gruenbaum:69,gruenbaum:74},
Winnie  \cite{winnie-70a,winnie-70b},
Malamet \cite{malamet},
Redhead \cite{redhead-93}
and Sarkar and Stachel \cite{stachel}
(cf. \cite{stanf-enc-syn} for a review).

In what follows, we shall frequently use  Einsteinian clocks based on ``radar coordinates.''
Thereby, we shall first fix an arbitrary unit distance.
Radar coordinate clocks use signals going back and forth two reflective
walls which are a predefined unit distance apart.
Time is measured by the number
of traversals of the signals between the walls.
Thus, if different signals are used to define time scales,
different time scales result.

In pursuing conventionalism further,
it appears not unreasonable to {\em assume}
the invariance of the speed of light as merely a convention
rather than as an empirical finding.
Indeed,
the International System
of standard units
\cite{speed-c} has implemented this approach.
Moreover,  the light cone structure already decides
the transformation of space-time coordinates:  Alexandrov's theorem
\cite{alex1,alex2,alex3,alex-col,robb,zeeman,heger,borchers-heger,goldblatt}
(cf. below; see \cite{benz,lester} for a review)
states that
the (affine) Lorentz transformations
are a consequence of the invariance of the speed of
light; a reasonable side assumption being
the one-to-one mapping of coordinates.

Introducing relativistic space-time transformations  as a
consequence
of conventions rather than of deep physics amounts to introducing relativity theory
``upside down,'' since in retrospect and in standard reviews
\cite{ein-05,ein1,urbantke} the Michelson-Morley and Kennedy-Thorndike
experiment is commonly presented as
an {\em experimental finding} supporting the assumption of the
invariance
of light in all reference frames.
Indeed, the very idea that the invariance of the two-way
velocity of
light is a mere convention might appear unacceptable. Yet, within a
given level of description, an unavoidable self-referentiality should be
acknowledged:
all experiments are themselves based upon coordinates (e.g., clocks
and scales) which operate with the very signals whose invariance is
experienced.

In contradistinction, the relativity principle, stating the {\em form invariance}
of the physical laws under such Lorentz transformations, conveys the
nontrivial
physical content. In this way, special relativity theory is effectively
split into a section dealing with geometric conventions and a different
one dealing with the representation of physical phenomena.

Thereby,  general covariance of the physical laws of motion can no
longer be required  globally. Indeed, form invariance will be satisfied
only relative to a  specific level of description; more precisely:
relative to a particular class of space-time transformations.

To give a simple
example:
Maxwell's equation are not form
invariant with respect
to Lorentz-type transformations generated from the assumption of the
invariance of the speed of sound; just as the description of
onedimensional sound phenomena propagating with velocity $\bar{c}$ by
$f(x-\bar{c}t)+g(x+\bar{c}t)$ is not invariant with respect to the usual Lorentz
transformations.
Yet, Maxwell's equation are form invariant with respect to the usual
Lorentz transformations; just as  $f(x-\bar{c}t)+g(x+\bar{c}t)$ is invariant
 with respect to
Lorentz-type transformations generated from the assumption of the
invariance of the speed of sound $\bar{c}$. [This can be checked by
insertion into equation
(\ref{e-ltt2}).]

With respect to a particular level of physical description, the time scale generated by the
corresponding signal may be
more appropriate than another if we adopt
Poincar\`e's criterion \cite{poinc06} resembling Occam's razor:
{\it``Time scales should be defined in such a way that the mechanical equations become as simple as possible.
In other words, there is no way to measure time which is more correct than another one; the one commonly
used is simply the most convenient.''}

This is a radical departure
from the requirement that the {\it fastest} signal  should be used for coordinatization. Of course,
today's fastest signal, light, is perfectly appropriate for today's fundamental description level of
elektromagnetism;
the corresponding scales (generated by the assumption of
invariance of the speed of light)
leaving the Maxwell
equations and other relativistic equations of motion
form invariant.
But that does by no means imply that different signals may not be more appropriate than light
for different levels of physical description.

\section{Lorentz-invariant media}

If, instead of light, sound waves
or water waves would be assumed constant in all inertial frames, then
very similar ``relativistic effects'' would result, but at a speed lower
than the speed of light.
This top-down approach to special relativity should be
compared to still another
bottom-up approach pursued, among others, by FitzGerald
\cite{FitzGerald},
J{\'{a}}nossy \cite{janossy},
Toffoli \cite{toffoli:79},
Erlichson \cite{erlich:73},
Bell \cite{bell-sr1,bell-92},
Mansouri and  Sexl
\cite{man-sexl:1,man-sexl:2,man-sexl:3},
Svozil
\cite{svo5,svo-86}, Shupe \cite{shupe} and
 G\"unther \cite{guenther}. There, relativistic forms are
derived from
``ether''-type theories.

Such considerations might well fit with speculations that the vacuum
might be a bifringent medium and that, for some unknown reason (e.g.,
the nonavailability of suitable detectors or the weakness of the
signal), only one of the two vacuum
``light'' cones has been observed so far.
As we shall mainly deal with the structual concepts of such findings, we
shall not discuss these asumptions further.

\section{Transformation laws}
%3
In what follows we shall consider the transformation properties of coordinates between
different reference frames; in particular between frames generated by two different
``maximal'' signalling speeds.
To be more precise, let $c$ denote the velocity of light.
{\em Alexandrov's theorem}
\cite{alex1,alex2,alex3,alex-col,robb,zeeman,heger,borchers-heger,goldblatt}
(cf. \cite{benz,lester} for a review),
states
that one-to-one mappings
$\varphi :{\Bbb R}^4\rightarrow {\Bbb R}^4$
preserving the Lorentz-Minkowski
distance {\it for light signals}
$$0=c^2 (t_x-t_y)^2-({\bf x}-{\bf y})^2=c^2 (t_x'-t_y')^2-({\bf x'}-{\bf
y'})^2,$$
$x=(t_x,{\bf x}),
y=(t_y,{\bf y})  \in {\Bbb R}^4$
are Lorentz transformations
$$x'=\varphi (x)=\alpha Lx+a$$
up to an affine scale factor $\alpha$.
(A generalization to ${\Bbb R}^n$ is straightforward.)
Hence, the Lorentz transformations appear to be essentially
derivable from the invariance of the speed of light alone.

Consider now that we assume the convention that, for one and the same
physical
system and for reasons not specified, another arbitrary but different
velocity $\bar{c}$ is invariant,
As a result of Alexandrov's theorem, a
different set of Lorentz-type transformation with $c$ substituted by
$\bar{c}$ is obtained. Of course, as can be expected, neither is
$\bar{c}$ invariant in the usual Lorentz frames, nor is $c$ invariant in
the Lorentz-type frames containing $\bar{c}$:
only the $c$-light cone appears invariant with respect to the
transformations containing $c$; the $\bar{c}$-light cone is not.
Conversely,
only the $\bar{c}$-light cone appears invariant with respect to the
transformations containing $\bar{c}$; the $c$-light cone is not.

Let us, for the sake of the argument, assume  that $c< \bar{c}$.
For all practical purposes, we shall consider two-way velocities
(measured back and forth).

As argued before, we shall consider two sets of inertial frames $\Sigma,
\bar{\Sigma}$ associated with $c$ and $\bar{c}$, respectively.
The set of all inertial frames $\Sigma $ is constructed by
 {\it a priori} and {\it ad hoc}  assuming that
$c$ is constant.
The set of all inertial frames $\bar{\Sigma} $ is constructed by
{\it a priori} and {\it ad hoc} assuming that
$\bar{c}$ is constant.

The construction of $\Sigma$ and $\bar{\Sigma}$ {\it via}
Alexandrov's principle
is quite standard.
Since $c$ and $\bar{c}$ are defined to be constant, two (affine) Lorentz
transformations
\begin{eqnarray}
&&x'=\varphi (x)=\alpha Lx+a
\quad
\textrm{ and }
\label{e-lt}
\\
\quad
&&\bar{x}' =\bar{\varphi } (\bar{x})=\bar{\alpha}
\bar{L}\bar{x}+\bar{a}
\label{e-ltt}
\end{eqnarray}
result for $\Sigma$ and $\bar{\Sigma}$, respectively. (In what follows, the
affine factors $\alpha ,\bar{\alpha}$ are set to unity.)
We shall also refer to these space and time scales as $c$-space,
$c$-time, and $\bar{c}$-space, $\bar{c}$-time, respectively.

The rules for constructing space-time diagrams for the twodimensional
problem (time and one space axis) are straightforward as well.
The Lorentz transformations
(\ref{e-lt}) and (\ref{e-ltt}) for $a=\bar{a}=0$ yield
\begin{eqnarray}
\varphi_v (x)=
(t',x_1',0,0)&=&
\gamma \left(    t-{vx_1\over c^2},x_1-vt,0,0 \right),
\textrm{ and }
\label{e-lt2}
\\
\bar{\varphi }_{\bar{v}} (\bar{x})=
(\bar{t}',\bar{x_1}',0,0)&=&
\bar{\gamma} \left(
\bar{t}-{\bar{v}\bar{x_1}\over
\bar{c}^2},\bar{x_1}-\bar{v}\bar{t},0,0 \right),
\label{e-ltt2}
\end{eqnarray}
with
$$
\gamma =+\left(1-{v^2\over c^2}\right)^{-{1\over 2}} \quad \textrm{ and
}\quad
\bar{\gamma} =+\left(1-{\bar{v}^2\over \bar{c}^2}\right)^{-{1\over 2}}
.$$
From now on, we shall write $x$ and $\bar{x}$ for $x_1$ and $\bar x_1$,
respectively. The second and third spatial coordinate will be omitted.

Consider faster-than-$c$ velocities $v$ in the range $$c<v\le \bar{c}.$$
For this velocity range,
 the Lorentz transformations
(\ref{e-lt2}), in
particular $\gamma$,
become imaginary  in the $\Sigma$-frames.  Therefore,
$\Sigma$ cannot account for such velocities.
For $\bar{\Sigma}$, these velocities are perfectly meaningful,
being smaller than or equal to $\bar{c}$.

The $x'$- and $t'$-axis is obtained by setting $t=0$ and $x=0$,
respectively. One obtains
\begin{equation}
{t}={{v}{x}\over {c}^2},\quad
\textrm{ and }              \quad
\bar{t}={\bar{v}\bar{x}\over \bar{c}^2},
\end{equation}
for the $x$- and $\bar{x}$-axis, as well as
\begin{equation}
t ={x\over v}, \quad
\textrm{ and }  \quad
\bar{t} ={\bar{x}\over \bar{v}},
\end{equation}
for the $t$- and $\bar{t}$-axis, respectively.

In general,
$c^2t^2-x^2\neq \bar{c}^2\bar{t}^2-\bar{x}^2$,
except for $c=\bar{c}$ and the coordinate frames cannot be directly
compared.
Thus the standard way of identifying unities does no longer work.

We might, nevertheless, generalize relativity theory by {\em requiring}
$
c^2t^2-x^2 = \bar{c}^2\bar{t}^2-\bar{x}^2.$
In this case, the identifications for unity are straightforward. In the
following, a different approach is pursued.

Another possibility is to proceed by constructing radar
coordinates in the following operational way.
Let us require that all frames
$\Sigma$     and
$\bar{\Sigma}$  have  the same origin. That is,
\begin{equation}
(t,x)=(0,0)\Leftrightarrow (\bar{t},\bar{x})=(0,0).
\end{equation}
Furthermore, let us consider the intrinsic coordinatization of two coordinate frames
$\sigma\in \Sigma$     and
$\bar{\sigma} \in \bar{\Sigma}$
 which are at rest with
respect to each other. As a consequence of the standard Einstein
synchronization conventions, two events which occur at the same
$c$-time in
$\sigma$  also
 occur at the same $\bar{c}$-time in
$\bar{\sigma}$.
Note that this concurrence of syncronicity is true only for the particular
frames $\sigma$ and $\bar{\sigma}$
and cannot be expected for all co-moving frames of $\Sigma$ and  $\bar{\Sigma}$.
At this point,
the preference of two frames $\sigma$, $\bar{\sigma}$
over others is purely conventional and does not reflect any ``deep
physics.''

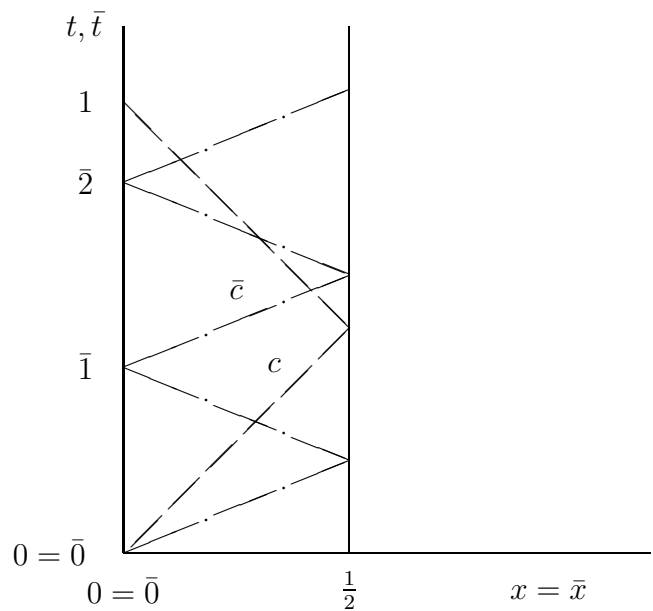
\begin{figure}
\begin{center}
%TexCad Options
%\grade{\off}
%\emlines{\off}
%\beziermacro{\off}
%\reduce{\on}
%\snapping{\off}
%\quality{2.00}
%\graddiff{0.01}
%\snapasp{1}
%\zoom{1.94}
\unitlength 1.00mm
\linethickness{0.4pt}
\begin{picture}(75.00,75.00)
\put(5.00,75.00){\line(0,-1){70.00}}
\put(5.00,5.00){\line(1,0){70.00}}
\put(35.00,5.00){\line(0,1){70.00}}
\put(5.00,5.00){\line(5,2){10.00}}
\put(17.00,10.00){\line(5,2){8.33}}
\put(27.33,14.23){\line(5,2){7.67}}
\put(26.33,13.67){\circle*{0.5}}
\put(16.00,9.33){\circle*{0.5}}
\put(5.00,29.67){\line(5,-2){10.00}}
\put(17.00,24.67){\line(5,-2){8.33}}
\put(27.33,20.44){\line(5,-2){7.67}}
\put(26.33,21.00){\circle*{0.5}}
\put(16.00,25.34){\circle*{0.5}}
\put(5.00,54.33){\line(5,-2){10.00}}
\put(17.00,49.33){\line(5,-2){8.33}}
\put(27.33,45.10){\line(5,-2){7.67}}
\put(26.33,45.66){\circle*{0.5}}
\put(16.00,50.00){\circle*{0.5}}
\put(5.00,29.67){\line(5,2){10.00}}
\put(17.00,34.67){\line(5,2){8.33}}
\put(27.33,38.90){\line(5,2){7.67}}
\put(26.33,38.34){\circle*{0.5}}
\put(16.00,34.00){\circle*{0.5}}
\put(5.00,54.33){\line(5,2){10.00}}
\put(17.00,59.33){\line(5,2){8.33}}
\put(27.33,63.56){\line(5,2){7.67}}
\put(26.33,63.00){\circle*{0.5}}
\put(16.00,58.66){\circle*{0.5}}
\put(35.05,34.93){\line(-1,-1){4.13}}
\put(30.06,29.94){\line(-1,-1){4.13}}
\put(25.08,24.96){\line(-1,-1){4.13}}
\put(20.09,19.97){\line(-1,-1){4.13}}
\put(15.11,14.99){\line(-1,-1){4.13}}
\put(10.12,10.00){\line(-1,-1){4.13}}
\put(35.05,34.93){\line(-1,1){4.13}}
\put(30.06,39.91){\line(-1,1){4.13}}
\put(25.08,44.90){\line(-1,1){4.13}}
\put(20.09,49.88){\line(-1,1){4.13}}
\put(15.11,54.87){\line(-1,1){4.13}}
\put(10.12,59.86){\line(-1,1){5.00}}
\put(-0.02,65.01){\makebox(0,0)[cc]{1}}
\put(-0.02,54.35){\makebox(0,0)[cc]{$\bar{2}$}}
\put(-0.02,29.77){\makebox(0,0)[cc]{$\bar{1}$}}
\put(-0.02,5.02){\makebox(0,0)[rc]{$0=\bar{0}$}}
\put(4.96,0.03){\makebox(0,0)[cc]{$0=\bar{0}$}}
\put(35.05,0.03){\makebox(0,0)[cc]{${1\over2}$}}
\put(-0.19,74.98){\makebox(0,0)[cc]{$t,\bar{t}$}}
\put(61.52,0.03){\makebox(0,0)[cc]{$x=\bar{x}$}}
\put(25.00,30.00){\makebox(0,0)[cc]{$c$}}
\put(20.00,40.00){\makebox(0,0)[cc]{$\bar{c}$}}
\end{picture}
\end{center}
\caption{\label{f-rcfl}
Construction of radar time coordinates. The vertical lines represent
mirrors for both signals at velocities $c$ (denoted by a dashed line
``$- - -$''), and
$\bar{c}$ (denoted by ``$-\cdot -\cdot -$'').}
\end{figure}

Let us first assume that we proceed by fixing one and the same unit of
distance for both coordinate systems; i.e.,
$x=\bar{x}$.
In such a case, the radar time coordinate $\bar{t}$ can be expressed in
terms of the radar time coordinate $t$ by
$\bar{t}=(\bar{c}/ c)t$.
This is illustrated in Figure~\ref{f-rcfl}.
In $c$-time $1$, the faster signal with velocity $\bar{c}$ has
been relayed back and forth the reflecting walls by a factor
$\bar{c}/c$.
Thus in summary, the transformation laws between $\sigma$ and $\bar{\sigma}$ are
\begin{equation}
(\bar{t}(t,x),\bar{x}(t,x))=({\bar{c}\over c}t,x).
\end{equation}

A more general conversion between
$\Sigma$     and
$\bar{\Sigma}$
involving moving coordinates is obtained by applying
successively the inverse Lorentz transformation
(\ref{e-lt2}) and
the Lorentz transformation
(\ref{e-ltt2}) with velocities $v$ and $\bar{w}$, respectively; i.e.,
\begin{equation}
(\bar{x}'',\bar{t}'')=
\bar{\varphi }_{\bar{w}} (\bar{t}',\bar{x}')=
\bar{\varphi }_{\bar{w}} ({\bar{c}\over c}t',x'),
\label{el-clt1}
\end{equation}
with
\begin{equation}
(t',x')={\varphi_v}^{-1}(t,x).
\label{el-clt2}
\end{equation}
More explicitly,
\begin{equation}
(\bar{t}''(t,x),\bar{x}''(t,x))=
\gamma \bar{\gamma} \left[
t\left({\bar{c}\over c}-{v\bar{w}\over \bar{c}^2}\right)
+
x\left({\bar{c}v\over c^3}-{\bar{w}\over {c}^2}\right),
t\left(v-{\bar{c}\bar{w}\over {c}}\right)
+
x\left(1-{\bar{c}v\bar{w}\over {c}^3}\right)
 \right]      .
\label{eq-lexltc}
\end{equation}
Here, $v<c$ and $\bar{w}<\bar{c}$.
As can be expected, for $c=\bar{c}$ and $v=\bar{w}$,
equation~(\ref{eq-lexltc}) reduces to
$
(\bar{t}''(t,x),\bar{x}''(t,x))=
(t,x)$.

Instead of identical space coordinates for two frames at rest with
respect to each other, we could have chosen
invariant time coordinates in both frames. A dual construction yields
the transformation laws
\begin{equation}
(\bar{t}(t,x),\bar{x}(t,x))=(t,{c\over \bar{c}}x).
\end{equation}

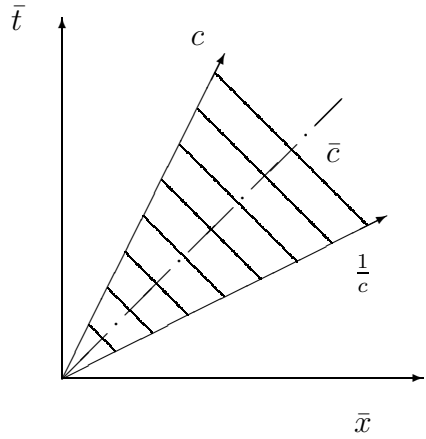
\begin{figure}
\begin{center}
%TexCad Options
%\grade{\on}
%\emlines{\off}
%\beziermacro{\off}
%\reduce{\on}
%\snapping{\off}
%\quality{6.00}
%\graddiff{0.01}
%\snapasp{1}
%\zoom{3.15}
\unitlength 1.20mm
\linethickness{0.4pt}
\begin{picture}(45.00,45.00)
\put(5.00,5.00){\vector(0,1){40.00}}
\put(5.00,5.00){\vector(1,0){40.00}}
\put(5.00,5.00){\vector(1,2){18.00}}
\put(5.00,5.00){\vector(2,1){36.00}}
\put(5.00,5.00){\line(1,1){5.00}}
\put(12.00,12.00){\line(1,1){5.00}}
\put(19.00,19.00){\line(1,1){5.00}}
\put(26.00,26.00){\line(1,1){5.00}}
\put(33.00,33.00){\line(1,1){3.00}}
\put(38.33,0.00){\makebox(0,0)[cc]{$\bar{x}$}}
\put(38.33,16.67){\makebox(0,0)[cc]{${1\over c}$}}
\put(35.00,30.00){\makebox(0,0)[cc]{$\bar{c}$}}
\put(20.00,42.67){\makebox(0,0)[cc]{$c$}}
\put(0.00,45.00){\makebox(0,0)[cc]{$\bar{t}$}}
\put(11.00,11.00){\circle*{0.4}}
\put(18.00,18.00){\circle*{0.4}}
\put(25.00,25.00){\circle*{0.4}}
\put(32.00,32.00){\circle*{0.4}}
%\emline(21.96,38.90)(38.87,21.98)
\multiput(21.96,38.90)(0.12,-0.12){141}{\line(0,-1){0.12}}
%\end
%\emline(19.95,34.88)(34.86,19.97)
\multiput(19.95,34.88)(0.12,-0.12){125}{\line(1,0){0.12}}
%\end
%\emline(18.05,30.97)(31.05,17.96)
\multiput(18.05,30.97)(0.12,-0.12){109}{\line(0,-1){0.12}}
%\end
%\emline(16.04,27.06)(27.03,15.96)
\multiput(16.04,27.06)(0.12,-0.12){92}{\line(0,-1){0.12}}
%\end
%\emline(14.03,23.04)(23.02,14.05)
\multiput(14.03,23.04)(0.12,-0.12){75}{\line(0,-1){0.12}}
%\end
%\emline(12.02,19.02)(19.00,12.04)
\multiput(12.02,19.02)(0.12,-0.12){59}{\line(0,-1){0.12}}
%\end
%\emline(10.01,15.00)(14.98,10.04)
\multiput(10.01,15.00)(0.12,-0.12){42}{\line(1,0){0.12}}
%\end
%\emline(8.00,10.99)(10.96,8.03)
\multiput(8.00,10.99)(0.12,-0.12){25}{\line(0,-1){0.12}}
%\end
\end{picture}
\end{center}
\caption{\label{fig-01}
Inertial frames of $\bar{\Sigma}$. The shaded area is forbidden for
frames of $\Sigma$ ($c<\bar{c}$).}
\end{figure}

Let us now consider  space-time diagrams for $\Sigma$ and
$\bar{\Sigma}$.
Figure \ref{fig-01} depicts twodimensional coordinate frames generated
for $\bar{\Sigma}$. The shaded region with the slope within $[c,{1\over
c}]$ are not allowed for $\Sigma$. They correspond to faster-than-$c$
frames.

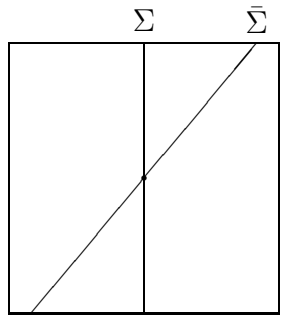
\begin{figure}
\begin{center}
%TexCad Options
%\grade{\off}
%\emlines{\off}
%\beziermacro{\on}
%\reduce{\on}
%\snapping{\off}
%\quality{2.00}
%\graddiff{0.01}
%\snapasp{1}
%\zoom{1.00}
\unitlength 0.6mm
\linethickness{0.4pt}
\begin{picture}(60.00,65.00)
\put(0.00,0.00){\line(1,0){60.00}}
\put(60.00,0.00){\line(0,1){60.00}}
\put(60.00,60.00){\line(-1,0){60.00}}
\put(0.00,60.00){\line(0,-1){60.00}}
\put(30.00,0.00){\line(0,1){60.00}}
\put(5.00,0.00){\line(5,6){50.00}}
\put(30.00,65.00){\makebox(0,0)[cc]{$\Sigma$}}
\put(55.00,65.00){\makebox(0,0)[cc]{$\bar{\Sigma}$}}
\put(30.00,30.00){\circle*{1.33}}
\end{picture}
\end{center}
\caption{\label{fig-01xy}
The set of all inertial frames of $\Sigma$ and $\bar{\Sigma}$ in the
set of all affine frames. The intersection between $\Sigma$ and
$\bar{\Sigma}$ represents frames of equal synchrony.}
\end{figure}

Figure \ref{fig-01xy} draws a reprentation of the sets of frames
$\Sigma$ and $\bar{\Sigma}$ in the set of all affine frames, denoted by
a square.

\section{Quasi time paradoxa and their resolution}

Since $\bar{c}> c$, superluminal signalling by any velocity $v$ with $\bar{c}
\ge v>c$ with respect to $c$ is an option for $\Sigma$. This could, at
least
from a straightforward point of view, result in quasi-time paradoxes,
such as Tolman's \cite{tolman,benf-book-new,recami:85,recami:87}
or G\"odel's paradoxes
\cite{godel-rmp,godel-sch,godel-ges3}. They originate from the fact
that, given superluminal signalling, signalling back in $c$-time is
conceivable, making a diagonalization argument
\cite{cantor-set1,cantor-set2,rogers1,odi:89}
similar to the classical
liar \cite{martin} possible \cite{svozil-paradox}.
Stated pointedly: given  free will,
this would enable an agent to send a signal
backwards in time if and only if the agent has not received this message
before.
Or, in a more violent version, kill the agent's own grandfather in
early childhood \cite{nahin}.
Likewise,  this would allow an agent to
become very knowledgeable, powerful and rich, which is not necessarily
paradoxical.

\begin{figure}
\begin{center}
$\;$\\
a)
%TexCad Options
%\grade{\on}
%\emlines{\off}
%\beziermacro{\off}
%\reduce{\on}
%\snapping{\off}
%\quality{6.00}
%\graddiff{0.01}
%\snapasp{1}
%\zoom{2.30}
\unitlength 0.90mm
\linethickness{0.4pt}
\begin{picture}(45.00,44.91)
\put(5.00,4.91){\vector(0,1){40.00}}
\put(5.00,4.91){\vector(1,0){40.00}}
\put(5.00,4.91){\line(1,1){5.00}}
\put(12.00,11.91){\line(1,1){5.00}}
\put(19.00,18.91){\line(1,1){5.00}}
\put(26.00,25.91){\line(1,1){5.00}}
\put(33.00,32.91){\line(1,1){3.00}}
\put(38.33,-0.09){\makebox(0,0)[cc]{${x}$}}
\put(35.00,29.91){\makebox(0,0)[cc]{${c}$}}
\put(0.00,44.91){\makebox(0,0)[cc]{${t}$}}
\put(41.49,10.13){\makebox(0,0)[cc]{$\bar{c}$}}
\put(12.06,6.44){\circle*{0.50}}
\put(12.49,29.88){\makebox(0,0)[cc]{$A$}}
\put(29.59,32.53){\makebox(0,0)[cc]{$B$}}
\put(15.32,37.06){\makebox(0,0)[cc]{$C$}}
%\vector(4.93,4.93)(42.75,19.86)
\put(42.75,19.86){\vector(3,1){0.2}}
\multiput(4.93,4.93)(0.30,0.12){125}{\line(1,0){0.30}}
%\end
%\vector(4.93,4.93)(20.00,43.05)
\put(20.00,43.05){\vector(1,3){0.2}}
\multiput(4.93,4.93)(0.12,0.30){126}{\line(0,1){0.30}}
%\end
%\emline(16.96,9.72)(32.03,43.05)
\multiput(16.96,9.72)(0.12,0.26){126}{\line(0,1){0.26}}
%\end
%\emline(5.07,4.93)(10.87,6.09)
\multiput(5.07,4.93)(0.58,0.12){10}{\line(1,0){0.58}}
%\end
\put(20.32,8.18){\circle*{0.50}}
\put(28.73,9.92){\circle*{0.50}}
\put(37.13,11.66){\circle*{0.50}}
%\emline(13.19,6.66)(18.99,7.82)
\multiput(13.19,6.66)(0.58,0.12){10}{\line(1,0){0.58}}
%\end
%\emline(21.59,8.40)(27.39,9.56)
\multiput(21.59,8.40)(0.58,0.12){10}{\line(1,0){0.58}}
%\end
%\emline(30.00,10.14)(35.80,11.30)
\multiput(30.00,10.14)(0.58,0.12){10}{\line(1,0){0.58}}
%\end
%\emline(38.40,11.88)(44.20,13.04)
\multiput(38.40,11.88)(0.58,0.12){10}{\line(1,0){0.58}}
%\end
\put(21.91,31.37){\circle*{0.50}}
%\emline(14.92,29.85)(20.72,31.01)
\multiput(14.92,29.85)(0.58,0.12){10}{\line(1,0){0.58}}
%\end
\put(21.91,33.56){\circle*{0.50}}
%\emline(22.90,31.59)(27.25,32.46)
\multiput(22.90,31.59)(0.54,0.11){8}{\line(1,0){0.54}}
%\end
%\emline(27.25,32.46)(22.75,33.48)
\multiput(27.25,32.46)(-0.50,0.11){9}{\line(-1,0){0.50}}
%\end
\put(42.32,16.67){\makebox(0,0)[cc]{$x'$}}
\put(16.09,42.90){\makebox(0,0)[cc]{$t'$}}
%\emline(20.87,33.91)(16.96,35.07)
\multiput(20.87,33.91)(-0.39,0.12){10}{\line(-1,0){0.39}}
%\end
\end{picture}
\\
$\;$\\
b)
%TexCad Options
%\grade{\on}
%\emlines{\off}
%\beziermacro{\off}
%\reduce{\on}
%\snapping{\off}
%\quality{6.00}
%\graddiff{0.01}
%\snapasp{1}
%\zoom{1.85}
\unitlength 0.90mm
\linethickness{0.4pt}
\begin{picture}(45.00,55.05)
\put(5.00,15.05){\vector(0,1){40.00}}
\put(5.00,15.05){\vector(1,0){40.00}}
\put(5.00,15.05){\line(1,1){5.00}}
\put(12.00,22.05){\line(1,1){5.00}}
\put(19.00,29.05){\line(1,1){5.00}}
\put(26.00,36.05){\line(1,1){5.00}}
\put(33.00,43.05){\line(1,1){3.00}}
\put(38.33,10.05){\makebox(0,0)[cc]{${x}'$}}
\put(35.00,40.05){\makebox(0,0)[cc]{${c}$}}
\put(0.00,55.05){\makebox(0,0)[cc]{${t}'$}}
\put(39.32,2.74){\makebox(0,0)[cc]{$\bar{c}$}}
\put(11.19,12.09){\circle*{0.50}}
%\emline(16.47,15.09)(16.47,55.01)
\put(16.47,15.09){\line(0,1){39.92}}
%\end
\put(1.76,39.01){\makebox(0,0)[cc]{$C$}}
\put(19.39,44.13){\makebox(0,0)[cc]{$B$}}
\put(7.64,51.84){\makebox(0,0)[cc]{$A$}}
%\emline(5.00,15.05)(10.19,12.64)
\multiput(5.00,15.05)(0.25,-0.11){21}{\line(1,0){0.25}}
%\end
\put(18.59,8.57){\circle*{0.50}}
\put(26.00,5.05){\circle*{0.50}}
\put(33.41,1.53){\circle*{0.50}}
%\emline(12.41,11.53)(17.59,9.12)
\multiput(12.41,11.53)(0.25,-0.11){21}{\line(1,0){0.25}}
%\end
%\emline(19.81,8.01)(25.00,5.60)
\multiput(19.81,8.01)(0.25,-0.11){21}{\line(1,0){0.25}}
%\end
%\emline(27.22,4.49)(32.41,2.08)
\multiput(27.22,4.49)(0.25,-0.11){21}{\line(1,0){0.25}}
%\end
%\emline(34.63,0.97)(39.81,-1.44)
\multiput(34.63,0.97)(0.25,-0.11){21}{\line(1,0){0.25}}
%\end
\put(11.19,47.22){\circle*{0.50}}
%\emline(5.00,50.18)(10.19,47.77)
\multiput(5.00,50.18)(0.25,-0.11){21}{\line(1,0){0.25}}
%\end
\put(11.19,42.25){\circle*{0.50}}
%\emline(5.00,39.30)(10.19,41.70)
\multiput(5.00,39.30)(0.25,0.11){21}{\line(1,0){0.25}}
%\end
%\emline(12.14,42.58)(16.29,44.74)
\multiput(12.14,42.58)(0.22,0.11){19}{\line(1,0){0.22}}
%\end
%\emline(16.29,44.74)(12.14,46.90)
\multiput(16.29,44.74)(-0.22,0.11){19}{\line(-1,0){0.22}}
%\end
\end{picture}
\\
$\;$\\
c)
%TexCad Options
%\grade{\on}
%\emlines{\off}
%\beziermacro{\off}
%\reduce{\on}
%\snapping{\off}
%\quality{6.00}
%\graddiff{0.01}
%\snapasp{1}
%\zoom{3.44}
\unitlength 0.90mm
\linethickness{0.4pt}
\begin{picture}(45.09,45.00)
\put(5.00,5.00){\vector(0,1){40.00}}
\put(5.00,5.00){\vector(1,0){40.00}}
\put(5.00,5.00){\line(1,1){5.00}}
\put(12.00,12.00){\line(1,1){5.00}}
\put(19.00,19.00){\line(1,1){5.00}}
\put(26.00,26.00){\line(1,1){5.00}}
\put(33.00,33.00){\line(1,1){3.00}}
\put(38.33,0.00){\makebox(0,0)[cc]{$\bar{x}$}}
\put(35.00,30.00){\makebox(0,0)[cc]{$\bar{c}$}}
\put(0.00,45.00){\makebox(0,0)[cc]{$\bar{t}$}}
\put(11.00,11.00){\circle*{0.50}}
\put(18.00,18.00){\circle*{0.50}}
\put(25.00,25.00){\circle*{0.50}}
\put(32.00,32.00){\circle*{0.50}}
%\vector(4.98,4.98)(10.02,44.89)
\put(10.02,44.89){\vector(1,4){0.2}}
\multiput(4.98,4.98)(0.12,0.95){42}{\line(0,1){0.95}}
%\end
%\vector(4.98,4.98)(45.09,10.02)
\put(45.09,10.02){\vector(4,1){0.2}}
\multiput(4.98,4.98)(0.95,0.12){42}{\line(1,0){0.95}}
%\end
%\emline(14.96,6.24)(20.00,44.89)
\multiput(14.96,6.24)(0.12,0.92){42}{\line(0,1){0.92}}
%\end
\put(6.94,20.02){\line(1,1){5.00}}
\put(12.94,26.02){\circle*{0.50}}
%\emline(13.99,26.97)(18.25,31.24)
\multiput(13.99,26.97)(0.12,0.12){36}{\line(1,0){0.12}}
%\end
\put(12.94,36.26){\circle*{0.50}}
%\emline(13.99,35.30)(18.25,31.04)
\multiput(13.99,35.30)(0.12,-0.12){36}{\line(1,0){0.12}}
%\end
%\emline(11.96,37.24)(9.44,39.76)
\multiput(11.96,37.24)(-0.12,0.12){21}{\line(-1,0){0.12}}
%\end
\put(8.18,18.45){\makebox(0,0)[cc]{$A$}}
\put(20.19,31.04){\makebox(0,0)[cc]{$B$}}
\put(11.38,40.53){\makebox(0,0)[cc]{$C$}}
\put(7.21,43.73){\makebox(0,0)[cc]{$\bar{t}'$}}
\put(43.25,7.40){\makebox(0,0)[cc]{$\bar{x}'$}}
%\emline(4.98,4.98)(9.15,11.09)
\multiput(4.98,4.98)(0.12,0.17){35}{\line(0,1){0.17}}
%\end
%\emline(10.21,12.73)(14.38,18.84)
\multiput(10.21,12.73)(0.12,0.17){35}{\line(0,1){0.17}}
%\end
%\emline(15.45,20.48)(19.61,26.59)
\multiput(15.45,20.48)(0.12,0.17){35}{\line(0,1){0.17}}
%\end
%\emline(20.68,28.23)(24.84,34.34)
\multiput(20.68,28.23)(0.12,0.17){35}{\line(0,1){0.17}}
%\end
%\emline(25.91,35.98)(30.07,42.08)
\multiput(25.91,35.98)(0.12,0.17){35}{\line(0,1){0.17}}
%\end
\put(29.69,38.31){\makebox(0,0)[cc]{$c$}}
\end{picture}
\end{center}
\caption{\label{f-cfrmg}
a)
Quasi-time paradox as perceived from $\Sigma$-frame $(t,x)$.
There is no apparent paradox here, because since $t_A<t_B<t_C$, no
information flows backward in time.
b)
Quasi-time paradox reveals itself when perceived from $\Sigma$-frame
$(t',x')$.
Information appears to flow backward in time, since
$t_A>t_B>t_C$.
c)
Resolution of the time paradox in
${\bar{\Sigma}}$-frame $(\bar{t},\bar{x})$.
In all $\bar{\Sigma}$-frames,
$\bar{t}_A<\bar{t}_B<\bar{t}_C$.           }
\end{figure}
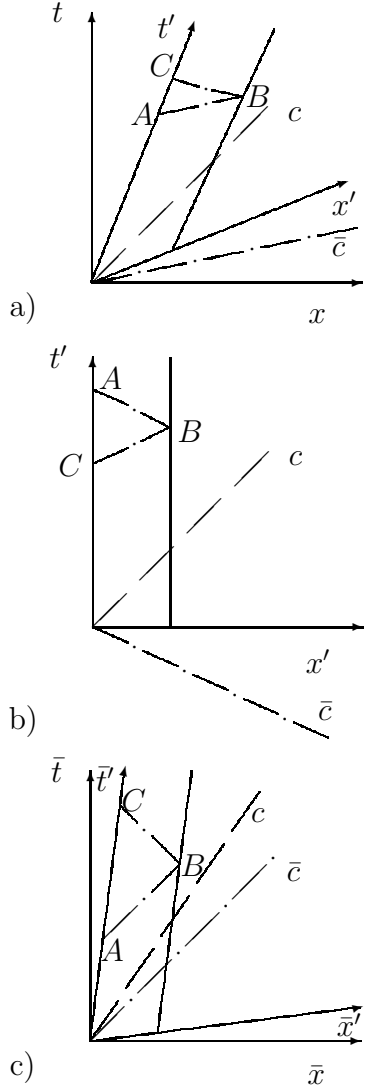

%5
To illustrate the quasi-paradoxical nature of the argument, let us
consider a concrete example. Assume as the two signalling speeds $c$ and
$\bar{c}$ the speed of sound and the speed of light, respectively.
Let us further assume that there exist intelligent
beings --- let us call them ``soundlanders'' --- capable of developing
physics in their
``ether''-medium \cite{shupe,svo5,svo-86,casti:96pr,guenther}.
For them, sound would appear as a perfectly appropriate phenomenon to
base their
coordinate frames upon. What if they discover sonoluminiscence; i.e.,
creation of signals at supersonar speeds $\bar{c}$? Surely, because of the
conceivable paradoxes discussed before,
this would result in a denial of the experimental findings at first and
in a crisis of (theoretical sound) physics later.
Fig. \ref{f-cfrmg} depicts the construction of a quasi-time paradox, as
perceived from the inertial frame $\Sigma$ generated by sound and the
inertial frame $\bar{\Sigma}$ generated by light.

%6
As can be expected, when viewed from $\bar{\Sigma}$, the  seemingly
``paradoxical'' process perceived by $\Sigma$ is not paradoxical at all.
It appears that
one resolution of the paradoxes is to switch the level of observation
and take the perspective that the ``true physics'' is not based upon
sound but on electromagnetic phenomena. After all, sound waves result
from
the coordinated motion of aggregates of atoms or molecules, which in
turn is dominated by the electromagnetic forces. In this extrinsic view,
the ``sound physics'' of the ``soundlanders'' is a representation of the
phenomena at an intermediate level of description
\cite{anderson:73,schweber:93}.
Since from that viewpoint, the appropriate signalling speed is
electromagnetic radiation at velocity $\bar{c}$, paradoxes disappear.
Thus any attempt to construct paradoxes at the intermediate
level of sound signals is
doomed to fail because that level of description turns out to be
inappropriate for the particular purpose.

%7
This extrinsic viewpoint is juxtaposed by the intrinsic viewpoint
\cite{bos,toffoli:79,svo5,svo-86,roessler-87,roessler-92,svozil-93}
of the
``soundlanders'' pretending to maintain their intermediate level of
``sound physics.'' For them, paradoxes are not realizable because
certain procedures or actions are not operational. This amounts to
the resolution of time paradoxes by the principle
of self-consistency \cite{friedetal} as   already
discussed, for instance, in Nahin's monograph \cite[p. 272]{nahin}.

\section{Concluding remarks}

As speculative as the above considerations may appear,
they can be brought forward consistently.
Even if exotic scenarios such as a birefringent vacuum appears highly unlikely, some
lessons for the presentation and interpretation of standard relativity
theory, in particular the splitting of conventions from the form
invariance of the physical laws, can be learned.

Theory --- in the
author's opinion for the worse --- tends to
exert a conservative influence in declining that faster-than-light or
  ``superluminal'' information communication and
 travel of the type
{\it ``breakfast on Earth, lunch on Alpha Centauri,
and home for dinner with your wife and children,
not your great-great-great grandchildren''} \cite{puthoff}
 is conceivable.
Accordingly,
any experimental, empirical claim of
allegedly superluminal phenomena  is confronted with the strongest
resistance from the theoretical orthodoxy, pretending on
the principal impossibility for superluminal communication.

The author is not convinced that as of
today there is reason to believe that there is experimental evidence of
faster-than-light
communication via tunnelling or other phenomena.
Yet, one cannot know when, if ever, superluminal
phenomena may be discovered. (In the author's opinion these would most
probably show up in an allegedly
nonpreservation of energy and momentum; very much in the same way as sonoluminiscence may be
viewed from the description level of sound.)
Hence, one purpose of this study has been the attempt to free experiment
from the pressure of the theoretical orthodoxy.
``Superluminal'' signalling {\it per se} is accountable for and does not
necessarily imply
``phenomenologic'' inconsistency.

From a system theoretic point of view, a
generalized principle of overall consistency of the
phenomena might be used to demontrate that too powerful agents would become inconsistent.
As a consequence, the predictive power as well as the
physical operationalizability (command over the phenomena) is limited by
this consistency requirement.
Events which may appear
undecidable and uncontrollable to an intrinsic observer bound by
incomplete knowledge may be perfectly controllable and decidable with
respect to a more complete theory.
In such a framework, different
signalling
speeds, in particular also superluminal signalling, can well be
accommodated  within a generalized theory of relativity. They do not
necessarily
mean inconsistencies but just refer to different levels of physical
descriptions and conventions which have to be careful accounted for.

\subsection*{Acknowledgements} The author is grateful for discussions
with and suggestions and comments by John Casti,
Georg Franck,
G\"unther Krenn,
Johann Summhammer,
Otto R\"ossler,
Franz-G\"unter Winkler
and
Helmuth Urbantke.
However, almost needless to say, their opinions do not necessarily
coincide with the approach expressed in this article;
nor should they be blamed for any misconception and fallacy of the author.

%\clearpage

%\bibliography{svozil}
%\bibliographystyle{aalpha}
%\bibliographystyle{plain}

\end{document}